\documentstyle[epsfig]{mn2e}

\def\be{\begin{equation}}
\def\ee{\end{equation}}
\def\bea{\begin{eqnarray}}
\def\eea{\end{eqnarray}}

\title[]{High Energy Afterglow Emission from Giant Flares of Soft
Gamma-Ray Repeaters: The Case of the 2004 December 27 Event from SGR
1806-20}

\author[]{Y. Z. Fan$^{1,2,3 \star}$, Bing Zhang$^{3 \star}$ and
 D. M. Wei$^{1,2}$
\thanks{E-mail: yzfan@pmo.ac.cn(YZF); bzhang@physics.unlv.edu(BZ); 
dmwei@pmo.ac.cn(DMW)} \\
$^1${\sl Purple Mountain Observatory, Chinese Academy of
Science, Nanjing 210008, China}\\
$^2${\sl National Astronomical
Observatories, Chinese Academy of Sciences, Beijing 100012,
China}\\
$^3${\sl Dept. of Physics, University of Nevada, Las Vegas, NV
89154, USA.}}

\date{Accepted ......  Received ......; in original form ......}

\begin{document}

\maketitle
\begin{abstract}
We discuss the high enegry afterglow emission (including high energy
photons, neutrinos and cosmic rays) following the 2004 December 27
Giant Flare from SGR 1806-20. If the initial outflow is relativistic
with a bulk Lorentz factor $\Gamma_0\sim {\rm tens}$, 
the high-energy tail of the synchrotron emission from electrons in the
forward shock region gives rise to a prominent sub-GeV emission, if the
electron spectrum is hard enough and if the intial Lorentz factor is
high enough. This signal could serve as a diagnosis of the initial
Lorentz factor of the giant flare outflow. 
This component is potentially detectable by GLAST if a similar giant
flare occurs in the GLAST era. With the available 10 MeV data, we
constrain that $\Gamma_0 < 50$ if the electron distribution is a
single power law. For a broken power law distribution of electrons, a
higher $\Gamma_0$ is allowed. 
At energies higher than 1 GeV, the flux is lower because of a high
energy cut off of the synchrotron emission component. The synchrotron
self-Compton emission component and the inverse Compton scattering
component off the photons in the giant flare oscillation tail are also
considered, but they are found not significant given a moderate
$\Gamma_0$ (e.g. $\leq 10$). The forward shock also accelerates cosmic
rays to the maximum energy $10^{17}$eV, and generate neutrinos with a
typical energy $10^{14}$eV through photomeson interaction with the
X-ray tail photons. However, they are too weak to be detectable. 
\end{abstract}


\section{Introduction}
The Soft Gamma-ray Repeater (SGR) 1806-20 lies in the Galactic plane,
at a distance of about $D_{\rm L}\approx 15.1$kpc (Corbel \&
Eikenberry 2004; cf. Cameron et al. 2005). A giant flare originated
from it on 2004 Dec. 27 is the brightest extra-solar transient event
ever recorded (e.g., Hurley et al. 2005; Palmer et al. 2005). Radio
follow-ups have resulted in detections of its afterglow (e.g., Cameron
et al. 2005; Gaensler et al. 2005).
Thanks to its brightness, an amazing variety of the data, including
the source size, shape, polarization and flux at multi-frequencies as
a function of time, have been collected (e.g., Gaensler et 
al. 2005; Cameron et al. 2005; Gelfand et al. 2005). Even so, our
understanding of the outflow is still in 
dispute. For example, the earliest afterglow data obtained so far is
about 7 days after the Giant flare. At this epoch, even an initially
relativistic outflow has been decelerated to the Newtonian phase by
the interstellar medium (ISM). As a result, whether the outflow is
relativistic initially (e.g. Wang et al. 2005; Dai et al. 2005) or not
(e.g. Gelfand et al. 2005; Granot et al. 2005) is uncertain. In
principle, similiar to the Gamma-ray Burst case (e.g., Krolik \& Pier
1991),  if the spectrum of the giant flare is nonthermal, a lower
bound of the initial Lorentz factor $\Gamma_0\sim {\rm tens}$ can be
derived from the so-called ``compactness argument'' (e.g., Huang et
al. 1998; Thompson \& Duncan 2001; Nakar, Piran \& Sari 2005; Ioka et
al. 2005). Observationally, the giant flare spectrum may be thermal
(Hurley et al. 2005) or nonthermal (Palmer et al. 2005), so that
$\Gamma_0$ could not be constrained well. 

In order to understand the dynamical evolution of the outflow better,
early multi-wavelength (including optical and hard $\gamma-$ray band)
observations are highly needed. The early optical emission has already
been calculated in Cheng \& Wang (2003) and Wang et al. (2005). In this
work, we focus on the high energy {\em afterglow} emission, including
sub-GeV photons (see \S{\ref{GeVP}}), high energy neutrinos and cosimic
rays (see \S{\ref{CosNeu}}).  High energy neutrinos from
magnetars in the quiescent state have been discussed by Zhang et
al. (2003). Assuming the internal shock mechanism, the neutrino,
cosmic ray and TeV photon emission accompanying the prompt 
giant flare have been discussed recently (Ioka et al. 2005;
Asano et al. 2005; Halzen et al. 2005).  

\section{High energy photon emission}\label{GeVP}
We first take $\Gamma_0=10$ as the typical Lorentz factor of the flow
to do sample calculations. The effect of varying $\Gamma_0$ will be
discussed later. The isotropic energy
of the outflow is taken as $E_{\rm iso}\sim 10^{46}$ergs. 
In the following analytical discussion, we assume that the shocked
electrons distributes as a single power-law $dn/d\gamma_{\rm e}\propto
\gamma_{\rm e}^{\rm -p}$ for $\gamma_{\rm m}<\gamma_{\rm
e}<\gamma_{\rm M}$, where $p\sim 2.5$, $\gamma_{\rm M}\sim
10^8{B'}^{-1/2}$ ($B'$ is the shock generted magnetic field strength,
see equation (\ref{DB})). Wang et al. (2005) find that a broken
power-law distribution of electrons, i.e., $dn/d\gamma_{\rm e}\propto
\gamma_{\rm e}^{\rm -p_1}$ for $\gamma_{\rm m}<\gamma_{\rm
e}<\gamma_{\rm b}$ and $dn/d\gamma_{\rm e}\propto \gamma_{\rm e}^{\rm
-p_2}$ for $\gamma_{\rm b}<\gamma_{\rm e}<\gamma_{\rm M}$, is required
to interpret the chromatic radio afterglow lightcurve steepening
around day 9. 
We therefore also include such a possibility in the numerical
calculations (see \S{\ref{Num}}).

With the standard parameters, the relativistic outflow is
decelerated by the ISM in a timescale 
\begin{equation}
t_{\rm dec}\approx 300{\rm s}~ E_{\rm
iso,46}^{1/3}n_0^{-1/3}\Gamma_{\rm 0,1}^{-8/3}, 
\label{t_dec}
\end{equation}
after which the ejecta moves with the Lorentz factor (for
$\Gamma>1/\theta_{\rm j}$)
\begin{equation}
\Gamma\approx 5.8E_{\rm iso,46}^{1/8}n_0^{-1/8}t_{\rm obs,3}^{-3/8}.
\label{Gamma_1}
\end{equation} 
where $n$ is the number density of the ISM, $t_{\rm obs}$ is the
observer time in unit of seconds. Throughout the work, we adopt the
convention $Q_{\rm a}=Q/10^{\rm a}$ using cgs units.  

As usual, we assume $\epsilon_{\rm e}$ and $\epsilon_{\rm B}$ as the
shock energy equipartition parameters for the shock accelerated
electrons and the magnetic fields, repsectively. The minimum electron
Lorentz factor reads
\begin{equation}
\gamma_{\rm m}\approx 184C_{\rm p}\epsilon_{\rm e,-0.5}(\Gamma-1).
\label{gamma_m}
\end{equation}  
where $C_{\rm p}=3(p-2)/(p-1)$. The strength of shock generated
magntic fields can be estimated as
\begin{equation}
B'\approx 3.9\times 10^{-2}{\rm Gauss}~\epsilon_{\rm
B,-2}^{1/2}n_0^{1/2}[\Gamma(\Gamma-1)]^{1/2}. 
\label{DB}
\end{equation}
Throughout the work, the superscript $'$  represents the parameter
measured in the comoving frame of the ejecta. 

\subsection{Inverse Compton Radiation}
A soft thermal X-ray tail emission modulated by the magnetar period is
typically detected after a giant flare hard spike. For the Dec. 27
event from SGR 1806-20, such a tail lasts for $T_{\rm tail}\sim 300$ 
with a typical photon energy
$\epsilon_{\rm X}\sim 30$keV (e.g. Mazets et al. 2005) and a
luminosity 
$L_{\rm X}\sim 2\times 10^{43} {\rm ergs~s^{-1}}~(t_{\rm obs}/50{\rm
s})^{-1}$. For $t_{\rm obs}<T_{\rm trail}$, besides the synchrotron
and synchrotron-self-Compton cooling processes (see \ref{SSC} for
detail), the electrons in the shocked region 
are also cooled by inverse Compton (IC) scattering off these X-ray tail
photons.    

Since $T_{\rm tail}$ is comparable to $t_{\rm dec}$, the ejecta has
not decelerated significantly, i.e., $\Gamma \sim \Gamma_0$. In the
comoving frame of the ejecta, the energy density of the X-ray tail
reads 
\begin{equation}
U'_{\rm X}\approx {L_{\rm X}\over 4\pi R^2 c \Gamma^2}\approx
0.27~{\rm ergs~cm^{-3}}~L_{\rm X,43}R_{\rm 15}^{-2}\Gamma_{1}^{-2}, 
\end{equation}
where $R$ is the radial distance of the forward-shock front from the
central source. 
On the other hand, the magnetic energy density generated in the
forward shock front reads 
\begin{equation}
U'_{\rm B}\approx 6\times 10^{-3}~{\rm ergs~cm^{-3}}~\epsilon_{\rm
B,-2}\Gamma_1^2n_0. 
\end{equation}  

In the rest frame of the shocked electrons with a random Lorentz
factor $\gamma_{\rm e}$, the energy of the thermal tail $\gamma_{\rm
e}\epsilon_{\rm X}/\Gamma$ is much larger than $m_{\rm e}c^2$, so that
the Klein-Nishina correction is important. For convience, we define
$x\equiv \gamma_{\rm e}\epsilon_{\rm X}/\Gamma m_{\rm e}c^2 \simeq
\gamma_{\rm e}/17\Gamma$. In the Klein-Nishina limit, $\sigma_{\rm
IC}=A(x)\sigma_{\rm T}$, 
where $A(x)\equiv {3\over 4}[{1+x\over x^3}\{{2x(1+x)\over 1+2x}-{\rm
ln}(1+2x)\}+{1\over 2x}{\rm ln}(1+2x)-{1+3x\over (1+2x)^2}]$, with the
asymptotic limits $A(x)\approx 1-2x+{26x^2\over 5}$ for $x \ll 1$, and
$A(x)\approx {3\over 8}x^{-1}({\rm ln}2x+{1\over 2})$ for $x \gg 1$
(e.g. Rybicki \& Lightman 1979). 

For illustration, we take $t_{\rm obs}=T_{\rm tail}$, at which
$R\approx 2\Gamma^2 cT_{\rm tail}\approx 1.8\times 10^{15}{\rm
cm}~\Gamma_1^2T_{\rm tail,2.5}$. For $\gamma_{\rm e}=\gamma_{\rm m}$,
we have $x=10.8$, $A(x=10.8)\approx 0.1$. The IC scattering is
therefore in the extreme Klein-Nishina limit, and the typical IC
photon energy can be well approximated by  
\begin{eqnarray}
h\nu_{\rm m}^{\rm IC}&\approx &h\Gamma\gamma_{\rm m} m_{\rm e}c^2\approx
\epsilon_{\rm e}[(p-2)/(p-1)](\Gamma-1)\Gamma m_{\rm p}c^2\nonumber\\ 
&\approx & 
9 {\rm GeV} ~ C_{\rm p}\epsilon_{\rm e,-0.5}\Gamma_1^2. 
\end{eqnarray}
The IC optical depth is
\begin{equation}
\tau\sim A(10.8)\sigma_{\rm T}n R/3\sim 4\times 10^{-11}n_0R_{15.26},
\end{equation}  
so that the 10 GeV photon luminosity can be estimated by
\bea
L_{\rm 10GeV} &\sim & \tau (L_{\rm X}/\epsilon_{\rm X})(h\nu_{\rm
m}^{\rm IC})\nonumber\\ 
&\sim & 4.3\times 10^{37}{\rm ergs~s^{-1}}~C_{\rm p}\epsilon_{\rm e,-0.5} n_0R_{15.26}t_{\rm
obs,2.5}^{-1}. 
\label{L10GeV}
\eea
where $h$ is the Planck constant.
For $T_{\rm tail}\leq t_{\rm dec}$, $\Gamma\sim \Gamma_0$, $R\propto
t_{\rm obs}$ we have $L_{\rm 10GeV}\propto Rt_{\rm obs}^{-1}\propto
t_{\rm obs}^0$. We can then estimate the total number of the photons
detectable by the {\em Gamma-Ray Large Area Telescope}
(GLAST)\footnote{http://glast.gsfc.nasa.gov/} in construction
\bea
N_{\rm tot}(10{\rm GeV})&\sim & [{A_{\rm GLAST}\over 4\pi D_{\rm
L}^2}\int_{0}^{\rm T_{\rm tail}}L_{\rm 10GeV}dt_{\rm obs}]/10{\rm
GeV}\nonumber\\ 
& \sim & 0.03~T_{\rm trail,2.5}n_0(D_{\rm L}/15.1{\rm kpc})^{-2},
\label{N_tot}
\eea
where $A_{\rm GLAST}\approx 8000{\rm cm^2}$ is the effective area
of the GLAST. Since usually at least 5 photons are needed to claim a
detection (e.g. Zhang \& M\'{e}sz\'{a}ros 2001 and references
therein), the above predicted $N_{\rm tot}$ is well below the
threshold of GLAST. 
This component is undetectable for an energetic giant flare 
similar to the recent one even for a much closer SGR, for example, SGR
1900+14.

The thermal tail photons would be also scattered by the electrons
accelerated by the reverse shock. 
The reverse shock is expected sub-relativistic. At $t_{\rm dec}$,
$\gamma_{34}\sim 1.2$, where $\gamma_{34}$ is the Lorentz factor of
shocked region relative to initial unshocked outflow. Therefore, for
the electrons accelerated by the reverse shock, one has $\gamma_{\rm
m}^{\rm r}=\epsilon_{\rm e}[(p-2)/(p-1)] (m_{\rm p}/m_{\rm e})
(\gamma_{34}-1)\sim 37$ by assuming the same parameters as in the
forward shock region.
Therefore $\epsilon_{\rm x}$ will be scattered to an energy $\sim
{\gamma_{\rm m}^{\rm r}}^2 \epsilon_{\rm x}\sim 30{\rm
MeV}$. According to Eqs.(\ref{L10GeV}) and (\ref{N_tot}), the detected
number of photons essentially depends on the IC optical depth $\tau$
and is independent on the typical energy of the photons. We can then
estimate the total number of the IC photons from the reverse shock
region by comparing that in the forward shock region.
First, the IC is now in the Thomson regime, i.e.
$\sigma_{\rm IC}\simeq \sigma_{\rm T}$. Second, the total number of
electrons contained in the reverse shock region is about $\Gamma_0$
times that in the forward shock region. The expected total number of
the 30 MeV photons is therefore
\begin{equation}
N_{\rm tot}({\rm 30 MeV})\sim \frac{\Gamma_0}{A(x=10.8)}
N_{\rm tot}({\rm 10GeV})\sim 3.
\label{N_tot2}
\end{equation} 
The actual value should be smaller since the timescale of having a
strong reverse shock could be shorter than $T_{\rm trail}$. Although
this $\sim 30$MeV reverse shock component is more prominent than the
$\sim 10$ GeV forward shock component, it is undetectable by GLAST,
either.
  
\subsection{Synchrotron and Synchrotron-self-Compton Radiation}\label{SSC}

For the forward shock emission, the cooling frequency
$\nu_{\rm c}$, the typical synchrotron frequency $\nu_{\rm
m}$ and the maximum spectral flux $F_{\rm \nu,max}$
read (e.g. Cheng \& Wang 2003; Wang et al. 2005)
\begin{equation}
\nu_{\rm c}=3.1\times 10^{19}{\rm
Hz}~E_{\rm iso,46}^{-1/2}\epsilon_{\rm B,-2}^{-3/2}n_0^{-1}{t}_{\rm
obs,3}^{-1/2}(1+Y)^{-2}, 
\label{nu_c}
\end{equation}  
\begin{equation}
\nu_{\rm m}=2.4\times 10^{12}{\rm
Hz}~C_{\rm p}^2E_{\rm iso,46}^{1\over 2}\epsilon_{\rm B,-2}^{1\over
2}\epsilon_{\rm e,-0.5}^2{t}_{\rm obs,3}^{-{3\over 2}},
\label{nu_m} 
\end{equation}
\begin{equation}
F_{\rm \nu,max}=474 {\rm Jy}~E_{\rm iso,46}\epsilon_{\rm
B,-2}^{1/2}n_0^{1/2}({D_{\rm L}\over 15.1{\rm kpc}})^{-2},
\label{Fmax}
\end{equation}
where $Y$ is the inverse Compton parameter, which can be estimated by
$Y\simeq [-1+\sqrt{1+4x\epsilon_{\rm e}/\epsilon_{\rm B}}]/2$ (e.g.,
Sari \& Esin 2001), where $x={\rm min}\{1,2.67(\gamma_{\rm
m}/\gamma_{\rm c})^{\rm (p-2)}\}$ is the radiation coefficient of the
shocked electrons (see equation (A8) of Fan, Zhang \& Wei
(2005a)), and $\gamma_{\rm c}$ is the electron cooling Lorentz factor
\begin{equation}
\gamma_{\rm c}\approx {7.7\times 10^{8}\over (1+Y)}{1\over \Gamma B'^2
t_{\rm obs}}. 
\label{gamma_c}
\end{equation}
Notice that only the synchrotron self-Compton is considered. The IC
component discussed in \S2.1 is in the extreme K-N regime at
$\gamma_{\rm c}$, giving a very small contribution to the $Y$
parameter. So it is neglected.
The resulting flux at a typical energy $h\nu_{\rm obs}=0.1$GeV can be
then estimated as
\begin{eqnarray}
F_{\nu_{\rm obs}}&=&F_{\rm \nu,max}~\nu_{\rm c}^{1/2}\nu_{\rm m}^{\rm
(p-1)/2}\nu_{\rm obs}^{\rm -p/2} 
=1.2\times 10^{-6}{\rm ergs~cm^{-2}}\nonumber\\
&& {\rm GeV^{-1}}~\epsilon_{\rm e,-0.5}^{\rm p-1}\epsilon_{\rm
B,-2}^{\rm p-2\over 4}E_{\rm 46}^{\rm {p+2\over 4}}C_{\rm p}^{\rm
(p-1)} ({D_{\rm L}\over 15.1{\rm kpc}})^{-2}
t_{\rm obs,3}^{\rm 2-3p\over 4}\nonumber\\
&& (1+Y)^{-1}({h\nu_{\rm obs} \over 0.1{\rm GeV}})^{\rm -(p-1)\over 2}.
\label{Fobs1}
\end{eqnarray}
For typical parameters, $Y$ is in order of 1, so the flux is much
higher than that of typical GRBs, so that this component can be
detected by GLAST. Notice that 
{\em there exists an upper limit on the synchrotron radiation
energy} 
\be
h\nu_{\rm M}\sim 2.8\times 10^6 h \gamma_{\rm M}^2\Gamma B'\sim
1.2 {\rm GeV} ~\Gamma_1,
\ee
above which a sharp cutoff is expected.

Since the outflow is only mildly relativistic, the collimation effect
is important in calculating the late lightcurves. Following Yamazaki
et al. (2005), we adopt a half-openning
angle of the collimated outflow $\theta_{\rm j}\approx
0.3$.
This leads to a geometry corrected energy $\approx  E_{\rm iso}\theta_{\rm
j}^2/4\approx 2.2\times 10^{44}{\rm ergs}~E_{\rm iso,46}(\theta_{\rm
j}/0.3)^2$, which matches that derived from the radio afterglow
modeling (e.g., Wang et al. 2005).
The above analytical calculations are only valid for
$\Gamma>1/\theta_{\rm j}$. For $\Gamma<1/\theta_{\rm j}$, the jet
sideways expansion effect is important. A rough estimate gives
$\Gamma(J_{\rm s})\propto t_{\rm obs}^{-1/2}$,  $F_{\rm
\nu,max}(J_{\rm s})\propto t_{\rm obs}^{-1}$, $\nu_{\rm c}(J_{\rm
s})\propto {\rm const.}$ and $\nu_{\rm m}(J_{\rm s})\propto t_{\rm
obs}^{-2}$ (e.g., Rhoads 1999), where $J_{\rm s}$ represents the jet
with important sideways expansion. Defining the ``jet break'' time
$t_{\rm j}\approx 4500{\rm s}~\Gamma_{\rm 0,1}^{-8/3}E_{\rm
iso,46}^{1/3}n_0^{-1/3}$, for $t_{\rm obs}>t_{\rm j}$, the flux could
be estimated as $F_{\nu_{\rm obs}}\simeq 1.8\times 10^{-7}{\rm
ergs~cm^{-2}~GeV^{-1}}~(t_{\rm obs}/t_{\rm j})^{\rm
-p}(1+Y)^{-1}(h\nu_{\rm obs}/0.1{\rm GeV})^{\rm -p/2}$. 
Since the typical Lorentz factor is small ($\sim 10$), the above
analytical treatment may not be a good approximation, and more
detailed numerical calculations are needed (see \S{\ref{Num}}).
The synchrotron-self-Component (SSC) luminosity ($L_{\rm SSC}$) could be
estimated through the $Y$ parameter, i.e. 
$Y={L_{\rm SSC}/ L_{\rm syn}}$. This results in the maximum SSC
spectral flux (for 
$t_{\rm dec}<t_{\rm obs}<t_{\rm j}$)
\begin{eqnarray}
F_{\nu_{\rm max}^{\rm SSC}}&\approx & Y \gamma_{\rm c}^{\rm
p-3}\gamma_{\rm m}^{\rm 1-p}F_{\rm \nu,max} 
\end{eqnarray}
where $\nu_{\rm m}^{\rm SSC}$ is the typical SSC frequency
\begin{eqnarray}
\nu_{\rm m}^{\rm SSC}&\approx & \gamma_{\rm m}^2\nu_{\rm m}
\approx 1.9\times 10^{18}{\rm Hz}~C_{\rm p}^4\epsilon_{\rm e,-0.3}^4\nonumber\\
&&\epsilon_{\rm B,-2}^{1\over 2}n_0^{-1/4}E_{\rm iso,46}^{3\over
4}t_{\rm obs,3}^{-9/4}. 
\end{eqnarray}
The resulting flux at $h\nu_{\rm obs}=0.1$GeV reads
\begin{eqnarray}
F_{\nu_{\rm obs}}&=&F_{\nu_{\rm max}^{\rm SSC}}~(\nu_{\rm obs}/\nu_{
\rm m}^{\rm SSC})^{\rm -(p-1)/2}\nonumber\\
&=&2.2\times 10^{-8}{\rm ergs~cm^{-2}~GeV^{-1}}~Y(1+Y)^{\rm p-3}C_{\rm
p}^{\rm p-1}\nonumber\\ 
&& \epsilon_{\rm e,-0.5}^{\rm p-1} \epsilon_{\rm B,-2}^{\rm
(13-3p)/4}n_0^{\rm (13-3p)/8}E_{\rm iso,46}^{\rm (15-p)/8}t_{\rm
obs,3}^{\rm (3-5p)/8}\nonumber\\ 
&&({D_{\rm L}/15.1{\rm kpc}})^{-2}(h\nu_{\rm obs}/0.1{\rm GeV})^{\rm
-(p-1)/2}. 
\label{Fobs2}
\end{eqnarray}
For $h\nu_{\rm obs}\leq 0.1$GeV, this radiation component is much
weaker than the synchrotron component. Beyond the synchrotron cutoff
at $h\nu_{\rm obs} \sim 1$GeV the SSC component dominates, but it is
well below the GLAST threshold.

\subsection{Numerical results}\label{Num}
\begin{figure}
\begin{picture}(-10,270)
\put(0,0){\includegraphics{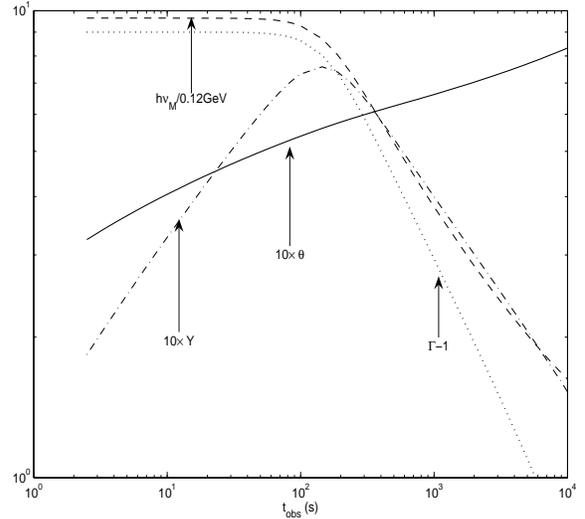}}
\end{picture}
\caption{Dynamical evolution of the ejecta. The dotted line represents
$\Gamma-1$; the solid line represents $\theta$
(multiplied by 10); the dash-dotted line represents
the SSC parameter $Y$ (multiplied by 10); and
the dashed line represents the cut off energy of synchrotron radiation
$h\nu_{\rm M}$ (normalized to 0.12GeV), all as functions of time.  
Following initial parameters are adopted. $E_{\rm iso}=10^{46}$ergs,
$\theta_{\rm j}=0.3$, $\Gamma_{\rm 0}=10$, $n=1{\rm cm^{-3}}$,
$\epsilon_{\rm e}=0.3$, $\epsilon_{\rm B}=0.01$ and $p=2.5$.  
The starting point of the calculation is taken as $R_0=10^{13}$cm.} 
\label{Dyn}
\end{figure}

\begin{figure}
\begin{picture}(-10,270)
\put(0,0){\includegraphics{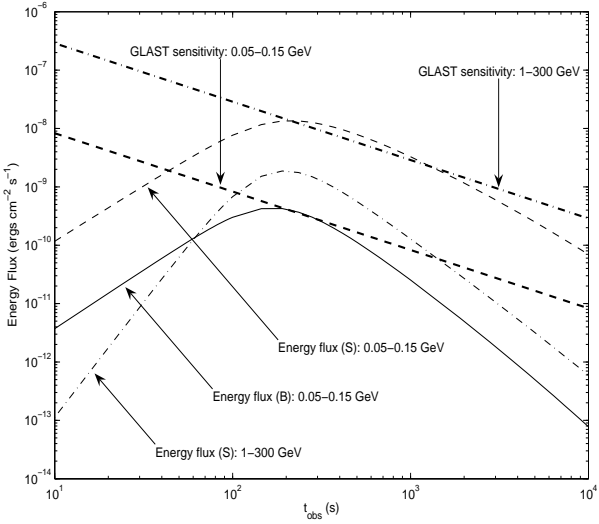}}
\end{picture}
\caption{Predicted photon energy fluxes as compared with the GLAST
flux sensitivity. The thin dashed line and the solid line represent
the flux in the energy range 0.05-0.15 GeV for different electron
energy distributions, where ``S'' denotes single power
law and ``B'' denotes broken power law. Only synchrotron component is
calculated since the SSC component is much dimmer. 
The thin dash-dotted line represents the flux in the 1-300 GeV energy
band. Only the SSC component is calculated since this is above the
synchrotron cutoff energy. 
The thick dashed line and the thick dash-dotted line represent the
GLAST threshold in the energy range 0.05-0.15 GeV and   
1-300 GeV, respectively. The GLAST threshold is defined by requiring
that during the integration timescale
$\sim t_{\rm obs}$ at least 5 photons are collected. For the
thin dashed and dash-dotted lines, same parameters as those
in Fig.\ref{Dyn} are taken. For the solid line, the parameters are
the same as those taken in Fig.\ref{Dyn} except that $p_1=2.2$,
$p_2=3.2$ (rather than $p$), and $\gamma_b / \gamma_m = 120$ are
adopted.}
\label{Rad}
\end{figure}

Similar to Huang et al. (2000) and Cheng \& Wang (2003), we have
calculated the dynamical evolution of the ejecta  (see Fig.\ref{Dyn})
and the accompanying high energy photon emission (see Fig.\ref{Rad})
numerically.  

As shown in Fig. \ref{Dyn}, the jet half-opening angle increases with
time rapidly. 
With sideways expansion,
the evolution of the jet half openning angle could be written as
(e.g. Huang et al. 2000) 
$d\theta/dt_{\rm obs}=c_{\rm s}(\Gamma+\sqrt{\Gamma^2-1})/R$, 
where 
$c_{\rm s}\approx \sqrt{(4\Gamma+3)(\Gamma^2-1)/
[3\Gamma(4\Gamma^2-1)]}c$ is the local sound speed. For $\Gamma \gg
1$, this could be approximated as $d\theta/dt_{\rm obs}\approx
1.2/(2\Gamma t_{\rm obs})$. It is apparent that the sideways expansion
of the jet is very important from the very beginning of the dynamical
evolution if the initial Lorentz factor is as small as 10. As a
result, there is no jet break in the $(\Gamma-1)$ lightcurve
(Fig.\ref{Dyn}) the energy flux lightcurve (Fig.\ref{Rad}).
This is different from the case of ultra-relativistic GRB outflows, in
which the sideways expansion is important only at later times.
One conclusion drawn from Fig.\ref{Dyn} is that the
ejecta accounting for the radio afterglow is nearly isotropic, which
matches the observations well (e.g. Cameron et al. 2005;
Galensler et al. 2005).   

According to Fig.\ref{Rad} where $\Gamma_0=10$ is adopted, we can see
that the predicted energy flux in the 0.05-0.15 GeV band is above the
GLAST sensitivity (thick dashed line), especially when a single
power-law electron energy distribution (thin dashed line) is adopted. 
If the electron distribution is a broken power law (solid line), the
detectability by GLAST is only marginal. 
In the energy band above 1 GeV, the predicted SSC energy flux (thin
dash-dotted line) is always below the GLAST sensitivity (thick
dash-dotted line), so that it is undetectable.

\begin{figure}
\begin{picture}(-10,270)
\put(0,0){\includegraphics{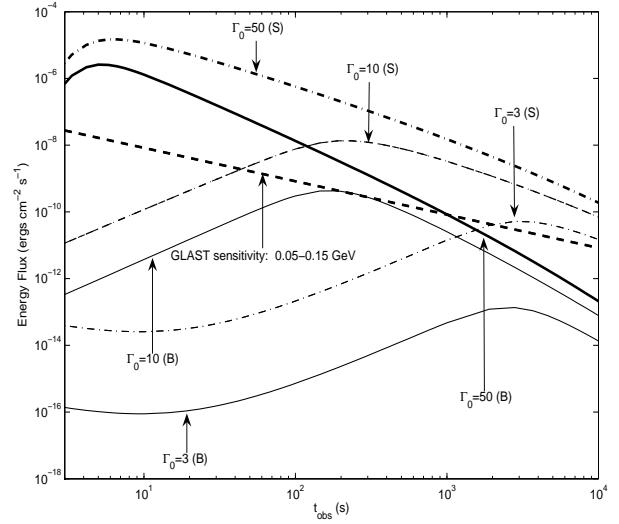}}
\end{picture}
\caption{The dependence of predicted 0.05-0.15 GeV photon energy flux
on the initial Lorentz factor $\Gamma_0$, whose actual values have been
marked in the figure. Both single power law (``S'') and broken power
law (``B'') distributions of the electrons are calculated. 
The thick dashed line represents the GLAST sensitivity in the
energy range 0.05-0.15 GeV. Except $\Gamma_0$, other parameters are
the same as those taken in Fig.\ref{Rad}.}  
\label{Rad2}
\end{figure}

In Fig.\ref{Rad2}, we investigate the dependence of the predicted
0.05-0.15 GeV energy flux (only the synchrotron radiation component
is taken into account) on $\Gamma_0$. The general trend is that a
higher $\Gamma_0$ leads to a stronger sub-GeV emission. For
$\Gamma_0\sim$ tens, regardless of the distribution of the
shocked electrons (single power law or broken power law), the
predicted fluxes are all above the GLAST. For $\Gamma_0\sim {\rm a
~few}$, only the single power law distribution model can yield to
marginally observable 0.05-0.15 GeV photon emission. 

In principle, a measurement of the sub-GeV flux in the GLAST era could
serve as a diagnosis of the intial Lorentz factor of the outflow.
For SGR 1806-60, available data already gives interesting constraints.
According to Matzets et al. (2005), the time averaged energy flux in
the $\epsilon_{\rm \gamma}\sim 10$ MeV band could be estimated as
$\epsilon_\gamma^2
dN/d\epsilon_\gamma \sim 1.6\times 10^{-6}{\rm ergs}~{\rm
cm^{-2}}~{\rm s^{-1}}$, where $dN/d\epsilon_\gamma
\sim 10^{-5}{\rm photons~cm^{-2}~s^{-1}~keV^{-1}}$ is the photon
number spectrum of the tail emission at 10MeV. Comparing with our
numerical results presented in Fig.\ref{Rad2}, the single power law
distribution model with $\Gamma_0=50$ is already above the observed
level\footnote{Our calculated energy flux is in the 
$\epsilon_\gamma=50-150$ MeV band. However, since
$\nu_{\rm obs} F_{\nu_{\rm obs}}\propto \nu_{\rm
obs}^{\rm (2-p)/2}$ very weakly depends on $\nu_{\rm obs}$, the
results could approximately apply to the 10 MeV band as well.}. 
A Lorentz factor $\Gamma_0 \geq 50$ is allowed only when a broken
power law distribution of the electrons is assumed.

\section{Cosmic rays and neutrinos}\label{CosNeu}

Below we estimate the maximum proton energy ($\epsilon_{\rm p}^{\rm
M}$) accelerated by the forward shock. For simplicity, we only discuss
$t_{\rm dec}<t_{\rm obs}<t_{\rm j}$.  In general, $\epsilon_{\rm
p}^{\rm M}={\rm min}[\epsilon_{\rm p}^{\rm M}(1), \epsilon_{\rm
p}^{\rm M}(2), \epsilon_{\rm p}^{\rm M}(3)]$ satisfies three
constraints (see also Fan, Zhang \& Wei 2005b). (1) The 
comoving shock acceleration time $t'_{\rm a}\sim \epsilon_{\rm
p}/\Gamma e B' c$ should be smaller than the comoving wind expansion
time $t'_{\rm d}\sim R/\Gamma c$, which yields $\epsilon_{\rm p}^{\rm
M}(1)\sim e B' R$. The numerical value reads
\be 
\epsilon_{\rm p}^{\rm M}(1)\simeq 2.4\times 10^{17}{\rm
eV}~\epsilon_{\rm B,-2}^{1/2}n_{0}^{1/8}E_{\rm iso,46}^{3/8}t_{\rm
obs,3}^{-1/8}.
\label{eps_p1}
\ee
(2) The comoving proton synchrotron cooling timescale $t'_{\rm
cool}=[6\pi m_{\rm p}^4c^3/\sigma_{\rm T}m_{\rm e}^2]\Gamma
\epsilon_{\rm p}^{-1} {B'}^{-2}$ should be longer than the comoving
acceleration timescale $t'_{\rm a}$, which
results in
\bea
\epsilon_{\rm p}^{\rm M}(2)\simeq  2.6\times 10^{21}{\rm eV}~
\epsilon_{\rm B,-2}^{-1/4}E_{\rm iso,46}^{1/16}
n_0^{-5/16}t_{\rm obs,3}^{-3/16}.
\label{eps_p2}
\eea
(3) The comoving proton cooling timescale due to photomeson
interaction should also be longer than the comoving acceleration
timescale $t'_{\rm a}$. However, from equation (\ref{nu_m}), the
typical frequency of the forward shock emission is too low to provide
the target photons for photomeson interactions at the $\Delta$
resonance, so the effect of photomeson interaction is too small to
change the proton cooling process.

Therefore, one has $\epsilon_{\rm p}^{\rm
M}=\epsilon_{\rm p}^{\rm M}(1)\sim 2.4\times 10^{17}$eV.
The source location of SGR 1806-20 is about $10^{\rm o}$ from the
Galactic center. In the region near the Galactic center the magnetic
field structure is poorly constrained. The time delay due to the
interstellar random magnetic field can be approximated as (e.g. Asano
et al. 2005) 
\begin{equation}
T_{\rm delay}\simeq (eB_{\rm G}D_{\rm L}/\epsilon_{\rm CR})^2(l/c),
\end{equation}
where $B_{\rm G}\sim 10^{-6}$G is the average magnetic field strength
in the Galaxy, $\epsilon_{\rm CR}$ is the typical cosmic ray energy,
and $l\sim 10-100$pc is the correlation length 
of the magnetic field (e.g., Asano et al. 2005). One then gets
$T_{\rm delay}\approx 6\times 10^5{\rm yr}~B_{\rm G,-6}^2(D_{\rm
L}/15.1{\rm kpc})^2(l/10{\rm pc})\epsilon_{\rm CR,17}^{-2}$. As a
result, these cosmic rays become a part of the cosmic ray background.    

As shown in equation (\ref{nu_m}), the typical frequency of the
forwardshock emission is too low to provide the target photons for  
the photomeson interaction at the $\Delta$-resonance. The only
interesting source of the neutrino emission is then the photomeson
interaction during the early epoch when the X-ray tail 
overlaps with the shocked region.
In the comoving frame of the ejecta, the thermal tail photons with
energy $\approx \epsilon_{\rm X}/\Gamma$ interact with the protons
with energy 
\be
\epsilon_{\rm p}\sim 0.3\Gamma^2{\rm GeV^2}/\epsilon_{\rm X}\simeq
10^{16}{\rm eV}~\Gamma_1^2(\epsilon_{\rm X}/30{\rm keV})^{-1}.
\ee
These protons lose $\sim 20\%$ of their energy at each $p\gamma$
interaction, dominated by the $\Delta$-resonance. Approximately, half
of pions are charged and decay into high energy neutrinos
$\pi^+\rightarrow \mu^++\nu_\mu\rightarrow e^++\nu_{\rm
e}+\bar{\nu}_\mu+\nu_\mu$, with the energy distributed roughly equally
among the decay products (e.g., Ioka et al. 2005). Therefore the
neutrino energy is $\sim 5\%$ of the proton energy, i.e., 
\be
\epsilon_{\rm \nu}\sim 5\times 10^{14}{\rm
eV}~\Gamma_1^2(\epsilon_{\rm X}/30{\rm keV})^{-1}. 
\ee   
The comoving number density of the thermal photons at the radius $R\sim
10^{15}$cm is 
\be
n_{\rm X}\approx U'_{\rm X}/(\epsilon_{\rm X}/\Gamma)\approx 5.5\times
10^7L_{\rm X,43} R_{15}^{-2}\Gamma_1^{-1}(\epsilon_{\rm X}/30{\rm
keV})^{-1}. 
\ee 
The fraction of the energy converted to pions can be estimated by the
number of the $p-\gamma$ interactions occuring within the shock with
the characteristic width $\Delta R\sim R/\Gamma$, i.e.  
\bea
f_\pi &\simeq &  0.2n_{\rm X}\sigma_\Delta R/\Gamma \nonumber\\
&\simeq  & 5.5\times 10^{-7}L_{\rm
X,43}R_{15}^{-1}\Gamma_1^{-2}(\epsilon_{\rm X}/30{\rm keV})^{-1}. 
\label{f_pi}
\eea
where $\sigma_\Delta\sim 5\times 10^{-28}{\rm cm^2}$ is the cross
section of the $\Delta-$resonance. For a neutrino detector with an
area $A_{\rm det}\sim 10^{10}{\rm cm^2}$, the expected event number
is
\be
N_{\rm \nu}\sim P_{\nu \rightarrow \mu} f_\pi A_{\rm det}E_{\rm
iso}/(32\pi D_{\rm L}^2\epsilon_\nu)\sim 7\times 10^{-5}, 
\label{N_event}
\ee
where $P_{\nu \rightarrow \mu}\simeq 3.5\times 10^{-4}(\epsilon_{\rm
\nu}/10^{15}{\rm eV})^{0.5}$ is the probability that a
neutrino produces a detectable high energy muon for $\epsilon_{\rm
\nu}>10^3{\rm TeV}$. We can see that the predicted neutrino number is
well below the detection threshold of the most powerful neutrino
detectors under construction. The main reason is that compared with
GRBs, $f_\pi$ (Eq.[\ref{f_pi}]) is much smaller.

\section{Summary}

We show that if a giant
flare similar to the 2004 Dec.27 event happens in the GLAST era, a
strong sub-GeV flare shortly after the flare (originated from the
hard tail of the synchrotron emission from the forward shock region)
should be detectable if the outflow is relativistic. A
positive/negative detection in the sub-GeV band would then give a
diagnosis of the initial Lorentz factor of the outflow. With the
available 10 MeV data (Mazets et al. 2005), we constrain $\Gamma_0 <
50$ for the Dec. 27 event if the electron distribution is a 
single power law, although a higher $\Gamma_0$ is allowed if the
electron distribution is a broken power law.
At higher energies (e.g. above 1 GeV), a cutoff of the synchrotron
emission is expected. Neither the synchrotron self-Compton emission in
the forward shock region, nor the inverse Compton off the X-ray tail
emission, could give a detectable flux for GLAST.

The forward shock is able to accelerate protons to an energy $\sim
10^{17}$eV. But the time delay for these cosmic rays to reach us is
very long, i.e. $\sim 10^6$ years. Neutrinos with an energy
$10^{14}$eV are also predicted, but the flux is too low to be
detected. Therefore, for a giant flare similar to the Dec. 27 event
taking place in the GLAST era, the most, and perhaps the only,
interesting high energy afterglow emission is the bright sub-GeV
photon emission lasting for thousands of seconds.   

\section*{Acknowledgments}
We thank  F. Halzen, Y. F. Huang and X. Y. Wang for
helpful comments. This work is supported by NASA NNG04GD51G and a NASA
Swift GI (Cycle 1) program (for B.Z.), the National Natural Science
Foundation (grants 10225314 and 10233010) of China, and the National
973 Project on Fundamental Researches of China (NKBRSF G19990754) (for
D.M.W.).


\begin{thebibliography}{99}
\bibitem[]{} Asano K., Yamazaki R., \& Sugiyama N. 2005
(astro-ph/0503335) 
\bibitem[]{} Cameron P. B. et al. 2005, Nature, 434, 1112
\bibitem[]{} Cheng K. S., \& Wang X. Y. 2003, ApJ, 593, L85
\bibitem[]{} Corbel S., \& Eikenberry S. S. 2004, A\&A, 419, 191
\bibitem[]{} Dai Z. G., Wu X. F., Wang X. Y., Huang Y. F. \&
Zhang B. 2005, ApJL, submitted
\bibitem[]{} Fan Y. Z., Zhang B., \& Wei D. M. 2005a, ApJ, in press
(astro-ph/0412105) 
\bibitem[]{} Fan Y. Z., Zhang B., \& Wei D. M. 2005b, ApJ, in press
(astro-ph/0504039)  
\bibitem[]{} Frail D., Kulkarni S. R., \& Bloom J. 1999, Nature,
398, 127 
\bibitem[]{} Gaensler B. M., et al. 2005, Nature, 434, 1104
\bibitem[]{} Gelfand J. D. et al. 2005, ApJ submitted
(astro-ph/0503269) 
\bibitem[]{} Granot J., et al. 2005, ApJ, submitted
(astro-ph/0503251) 
\bibitem[]{} Halzen F., Landsman H., \& Montaruli T. 2005
(astro-ph/0503348) 
\bibitem[]{} Huang Y. F., Dai Z. G., \& Lu T. 1998,
Chin. Phys. Lett., 15, 775 
\bibitem[]{} Huang Y. F., Gou L. J., Dai Z. G., \& Lu T. 2000, ApJ. 543, 90
\bibitem[]{} Hurley K., et al. 2005, Nature, 434, 1098
\bibitem[]{} Ioka K., Razzaque S., Kobayashi S., \& Meszaros
P. 2005, ApJL submitted (astro-ph/0503279) 
\bibitem[]{} Krolik J. H., \& Pier E. A.  1991, ApJ, 373, 277
\bibitem[]{} Rybicki G. B., \& Lightman A. P. 1979, Radiative
Processes in Astrophysics (New York: Wiley) 
\bibitem[]{} Mazets E. F., et al. 2005 (astro-ph/0502541)
\bibitem[]{} Nakar E., Piran T., \& Sari R. 2005 (astro-ph/0502052)
\bibitem[]{} Palmer D. A., et al. 2005, Nature, 434, 1107 
\bibitem[]{} Sari R., \& Esin A. A. 2001, ApJ, 548, 787
\bibitem[]{} Rhoads J. E. 1999, ApJ, 525, 737
\bibitem[]{} Thompson C., \& Duncan R. C. 2001, ApJ, 561, 980
\bibitem[]{} Wang X. Y., Wu X. F., Fan Y. Z., Dai Z. G., \& Zhang
B. 2005, ApJ, 623, L29 
\bibitem[]{} Yamazaki R., Ioka K., Takahara F., \& Shibazaki
N. 2005, PASJ in press (astro-ph/0502320) 
\bibitem[]{} Zhang B., Dai Z. G., M\'{e}sz\'{a}ros P., Waxman E.,
\& Harding, A. K. 2003, ApJ, 595, 346 
\bibitem[]{} Zhang B., \& M\'{e}sz\'{a}ros P. 2001, ApJ, 559, 110
\end{thebibliography}
\end{document}